\documentclass[10pt,journal]{IEEEtran}

\usepackage[utf8]{inputenc}
\usepackage{amsmath,amsfonts,amsbsy,amssymb}
\usepackage{mathabx}
\usepackage{mathrsfs}
\usepackage{tabularx}
\usepackage{graphicx}
\usepackage{cite}
\usepackage{wasysym}
\usepackage{multirow}
\usepackage{float}
\usepackage{color}
\usepackage[colorlinks=false,urlcolor=black,hidelinks]{hyperref}
\usepackage{epstopdf}

\begin{document}

\title{Energy Internet via Packetized Management: Enabling Technologies and Deployment Challenges}
\author{Pedro H. J. Nardelli, Hirley Alves, Antti Pinomaa, Sohail Wahid,\\ Mauricio C. Tomé, Antti Kosonen, Florian Kühnlenz, Ari Pouttu, Dick Carrillo
\thanks{P. H. J. Nardelli, Antti Pinomaa, Sohail Wahid, Antti Kosonen and Dick Carrillo are with Laboratory of Control Engineering and Digital Systems, Lappeenranta University of Technology, Finland. Hirley Alves, Mauricio C. Tomé, Florian Kühnlenz and Ari Pouttu are with the Centre for Wireless Communications (CWC) at University of Oulu, Finland. P. H. J. Nardelli is also with CWC. This work is partly funded by Academy of Finland via BCDC Energy (SRC/n.292854), ee-IoT (ICT2023/n.319009) and 6Genesis Flagship (n. 318927). Contact: pedro.nardelli@lut.fi}%
}
\maketitle
%
%
\begin{abstract}
This paper investigates the possibility of building the Energy Internet via a packetized  management of non-industrial loads.
The proposed solution is based on the cyber-physical implementation of energy packets where flexible loads send use requests to an energy server.
Based on the existing literature, we explain how and why this approach could scale up to interconnected micro-grids, also pointing out the challenges involved in relation to the physical deployment of electricity network.
We then assess how machine-type wireless communications, as part of 5G and beyond systems, will achieve the low latency and ultra reliability needed by the micro-grid protection while providing the massive coverage needed by the packetized management.
This more distributed grid organization also requires  localized governance models.
We cite few existing examples as local markets, energy communities and micro-operator that support such novel arrangements.
We close the paper by providing an overview of ongoing activities that support the proposed vision and possible ways to move forward.
\end{abstract}


\section{Introduction}
The last two decades have witnessed a revolutionary change on how people communicate.
Non-stop technological evolution of personal computers, Internet and mobile (cellular) networks converged upon the upcoming telecommunication standards (refer, for instance, to  5G-PPP use cases \cite{elayoubi20175g}).
The widespread of relatively cheap wireless-enabled sensors for data acquisition together with more effective algorithms for data processing makes the daily life increasingly digitalized.
New concepts are frequently arising in such a context.
Terms like Internet of Things (IoT), Big Data, Data Analytics and even Internet of Everything are constantly appearing.

Energy Internet is one of these terms; it causes impact at first but, after a careful examination, no clear definition can be found.
The majority of (high quality) scientific works refers to the Energy Internet as the next stage of the Smart Grid, which is also another unclear concept.
From our understanding, these two concepts relate to a more efficient management of energy systems -- usually focused on electricity grids -- based on information and communication technologies (ICTs).
In this sense, the large amount data (big data) generated during various process in the smart grid systems including power generation, energy storage and delivery create many possible advantages for a more intelligent system management \cite{wang2017wireless,wang2017distributed}, but also brings threats to its stability \cite{wang2017big}.

In this contribution, we rather want to answer the following question: \textit{what is Energy Internet and what makes it so different from other existing and potential solutions?} 
Our view is that Energy Internet is the cyber-physical system that virtualizes the management of the distribution grid via discretized packets.
This packetized energy management is inspired by the data internet and may be extended to the management of interconnected micro-grids based on energy routers \cite{abe2011digital}.

\begin{figure}[!t]
	\centering
	\includegraphics[width=1\columnwidth]{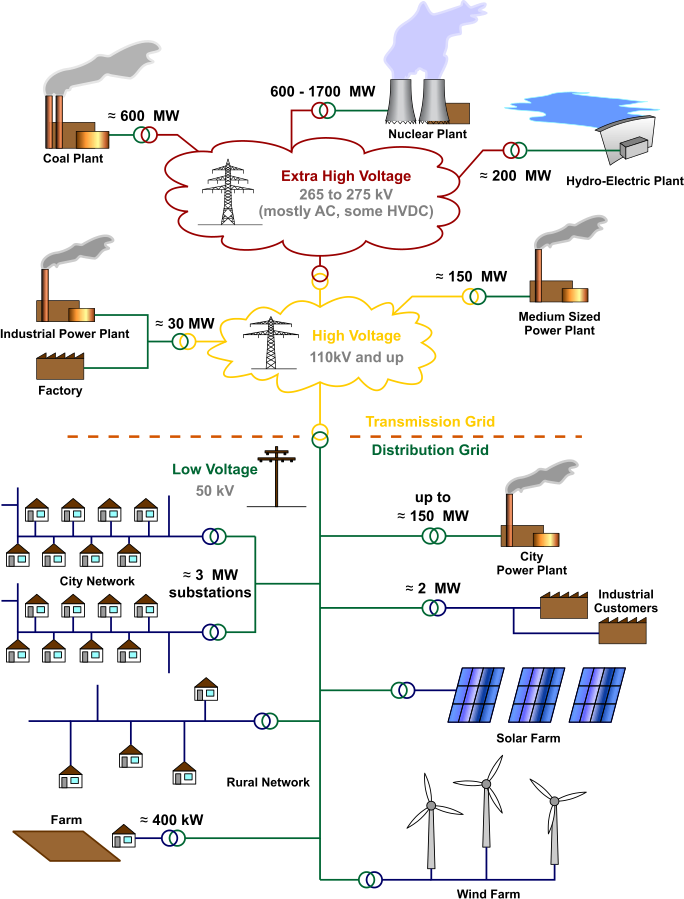}
	\caption{\label{fig:grid}Illustration of the electric power grid including large-scale power plants connected to the distribution grid and large-scale industries via high-voltage networks, and smaller scale solar and wind generation connected directly to the distribution grid. Source: \url{https://commons.wikimedia.org/wiki/File:Electricity_Grid_Schematic_English.svg}. By MBizon [CC BY 3.0]  \url{https://creativecommons.org/licenses/by/3.0}, from Wikimedia Commons.}
	%
\end{figure}

Roughly speeaking, energy is virtually split into packets to be consumed during a certain time period, e.g. $x$ watt-hour in $y$ minutes.
This concept is not new since packetized energy has been researched and implemented for some time either in a real physical packets (e.g. \cite{abe2011digital,Ma2018optimal,Ma2018elastic}) or virtualized ones (e.g. \cite{saitoh1996new,zhang2012packetized,almassalkhi2017packetized,almassalkhi2018asynchronous}), although it has never become mainstream.
What we are proposing here is a way that packetized energy management could scale up at the distribution grid level (Fig. \ref{fig:grid}) so it may become dominant.
This solution would be the basis of the demand-side management in a future dominated by micro-grids supplied by renewable sources.

This contribution points out that: (i) virtual energy packets is the most suitable way to manage the electricity inventory within micro-grids, (ii) machine-type wireless communication techniques are able to provide the ultra-reliability, low latency and massive connectivity needed to sustain the micro-grid applications (e.g. from protection to telemetry), and (iii) new governance arrangements based on local markets or energy communities are required to sustain the solution in the socio-economical domains.
In this sense, the proposed Energy Internet is now technically feasible due to the recent advances of ICTs, mainly in machine-type communications (MTC).
Moreover, it is timely in both political and social terms, as indicated by the strong pressure for more renewable sources, better energy efficiency, development of community-led energy management and the end of energy poverty.
Given the favorable conjuncture, this paper indicates how the Energy Internet could emerge given the recent research in packetized energy management, micro-grids, MTC and new governance models.

We present the proposed idea in the following way. 
We first review in Section \ref{sec:EnergyIntenet} the literature of packetized energy,  showing the differences between its physical and cyber-physical deployments.
Still in this section, we introduce the challenges involved when scaling up the cyber-physical deployment of the packetized energy management.
This includes from very small time-scales (milliseconds in grid protection mechanisms) to decades (long-term investments).
We then focus in Section \ref{sec:Applications} on the issues related to the electricity grid that supports the proposed cyber-physical deployment.
There we investigate the issues related to the physical grid, its modernization based on intelligent devices and decentralization tendency towards micro-grids, showing a simple study-case of how the proposed energy server would function.
Section \ref{sec:MTC} focused on the recent advances in wireless systems, mainly in machine-type communications (MTC), as an enabler technology capable of operating in different regimes, including ultra-reliable low-latency communications (grid protection) and massive connectivity (energy management/tertiary control).
As the proposed solution is based on decentralization of the distribution grid, it requires new governance models to be effective.
Section \ref{sec:other-issues} discusses new elements like localized energy markets and energy communities that would allow for scaling up the proposed packetized management.
Section \ref{sec:future} covers three ongoing research activities by the authors that support our vision presented.
We conclude the paper in Section \ref{sec:final} by reinforcing the feasibility of the proposed solution and indicating a possible transition plan.

\section{Energy Internet}
\label{sec:EnergyIntenet}

\subsection{Background}
Back in 2004, the prestigious weekly magazine \textit{The Economist} published an article called \textit{Building the energy internet} \cite{economist}.
The argument went along the idea of decentralization of production and how information and communication technologies can be used to build the Energy Internet -- a smarter, more intelligent, electricity grid.
After fourteen years of technological development and political struggles, some predictions have been indeed realized while others not.
We will present next some of the important advances following this line of thought.

For example, in \cite{tsoukalas2008smart}, the authors have discussed the idea of a virtualized buffer for demand-side management, where the actual consumption can be scheduled according to predictions in supply and demand.
Although motivated by an actual deployment of deregulated electricity markets, we argue that the article only provided hints and a rough road-map.
In \cite{huang2011future,su2013proposing}, the so called \textit{Future Renewable Electric
Energy Delivery and Management} (FREEDM) was discussed as the way to build the Energy Internet based on a deregulated energy market.
Their concept relies on a plug-and-play interface (like a USB),  an energy router (motivated by the internet router) so that \textit{on} and \textit{off} reference signals can be send to the connected devices, and open-standard-based operating system to coordinate actions.
In \cite{huang2011future}, the authors specifically dealt with the FREEDM cyber-physical implementation (from  solid-state research to distributed grid intelligence).
In \cite{su2013proposing}, they looked at possible market arrangements based on game theory to support such approach.

{More recently, Gao et al. proposed in \cite{GaoTIE2018} a verification solution for an energy router based systems. 
A continuous-time Markov chains mode is used when the system has multiple routers to check the reliability of system operation. 
The electricity trading was taken as an example to verify the real world energy router based system functions. 
A Markov decision process was used to check the trading behavior. 
For system performance, an energy scheduling process was implemented on cloud computing experimental tool, verifying the effectiveness of the   proposed schemes.}

{In \cite{wang2018survey}, Wang et al. discussed about the evolution of smart grid to their proposed Energy Internet, following the architecture of FREEDM system.
They summarized the necessary conditions and requirements that the Energy Internet has to satisfy to enable the optimal use of highly scalable and distributed energy resources. 
By comparing smart grids with the proposed solution, they indicated various advantages of the proposed system in relation to  reliability and security, enabling safe transfer of energy units to the customers. }
There is in fact an extensive literature in the topic, mostly addressing how to optimize the energy management (either in distributed or decentralized fashion).
These works are quite important, but they are not about Energy Internet as we describe it here (although they are certainly part of it).
Such contributions have never fully considered that energy can be managed via ``discretized'' energy packets; \textit{The Economist} put it plainly: \cite{economist}:
``Of course, the power grid will never quite become the internet -- it is impossible to packet-switch power.''
The present article -- supported by previous research in power systems and communications engineering  -- disagrees with such a claim; treating energy as packets is not only technically sound and socioeconomically feasible but is rather the key building block of the  Energy Internet.

\subsection{The beginning of energy-as-packets}
The idea of approaching energy networks using discretized energy packets, similar to the way data networks and the internet is managed, is not new.
Back in 1996, two Japanese researchers, Saitoh and Toyoda, introduced in \cite{saitoh1996new} the concepts of ``open electric energy network'' and  ``packet electric power transportation.''
The first indicates that the energy network shall be opened to small, distributed, generators and storage devices.
The second refers to the way the excess energy is managed in such a network.
The authors proposed a conceptual system where this excess is packetized and their usage depends on the demand patterns from the consumers.
Specifically, the authors dimensioned the storage capacity of the system based on queuing theory, as in data networks.
Deploying such a solution, however, would require ICTs as well as storage capabilities neither affordable nor available at that time.
In this sense, we can say that the authors were visionaries.

\subsection{Physical packets and the digital grid}
\label{subsec:phy}

More than a decade later, Abe et al. proposed in \cite{abe2011digital} the concept of  \textit{Digital Grid} in which the large connected grid shall be divided into smaller grids, which works asynchronously and are connected via digital grid routers.
These routers communicate for sending power between the segmented grids using the same power lines via a direct current (DC) and \textit{physical energy packets}.
Using Internet Protocol (IP) inspired addresses and power line communication (PLC), the authors claim that proposed architecture would lead to a more efficient use of power lines and accommodate distributed sources, allowing new digitalized services and accounting.

In \cite{takahashi2015router}, the authors further developed the results from Abe et al. by including a router-to-router power transfer.
It also uses the storage capabilities from the routers to allow tunable starting time.
This would build a \textit{networked power packet distribution system}.
In a series of papers (e.g. \cite{gelenbe2012energy,gelenbe2016energy}), Gelenbe and his team also developed the idea of \textit{energy packet networks}, but relying on the G-Network theory \cite{gelenbe1994g} (a generalization of queuing theory that incorporates, for instance, traffic re-routing or traffic destruction).
This line of work is, however, very abstract so the actual constraints from the power grid are poorly addressed.
In a more realistic environment, the authors in \cite{rodriguez2015experimental} validated their proposed \textit{intelligent power router} using an experimental test-bench and simulations.

In very recent contributions \cite{Ma2018optimal,Ma2018elastic}, the authors proposed a distribution grid with local area packetized power networks.
In \cite{Ma2018elastic}, the authors investigated the integration of distribution grid with local area packetized power networks by proposing a multi-mode energy distribution system.
The system would work in phases as follows:
A storing phase during the off-peak hours and consumption phase for fulfilling the load fluctuations in real time.
In this scheme, coordination between alternate current (AC) and DC sectors is done while the elasticity of the DC loads are
checked to reduce the load fluctuations.
This scheme requires on its turn a reliable and massive communication network to
control and monitor all operations, management and process related to the energy exchange.

In \cite{Ma2018optimal}, the authors considered a local area
packetized power network where all the subscribers are linked by multi-channel
power routers.
A special protocol for and efficient transmission and subscriber matching is
proposed for regulation of its operations.
In this process the subscribers are matched with the demand suppliers and then the routers helps to transmit power over its different channels.
Based on the matching theory, a subscriber matching phenomenon is
introduced to increase the benefits for each subscriber and create and algorithm to solve the
problem.
Also a heuristic solution for the transmission scheduling was introduced to keep a balance within the network.
The simulation results confirmed the effectiveness of the proposed
method in increasing the subscribers benefits and producing fairness of power line occupation between the subscribers.

\subsection{Cyber-physical packets}

Using a similar terminology another line of research consists of treating AC power consumption as discretized packets as in \cite{tsoukalas2008smart}; in this case we can say that the energy packets become cyber-physical.
In this scenario, specific appliances have their consumption discretized thus the load management  and control can be performed to achieve a pre-determined goal.
We can cite \cite{lee2011demand} where Lee et al.  proposed a technique for load management of air conditioner in large apartment complexes based on queuing theory and Markov birth-death processes.
This method was shown to be effective for managing loads in respect to both customer and power companies. 
Interestingly, the load management and aggregation are performed by \textit{tokens} (related to energy packets), which are issued and allocated for giving rights to the companies to turn on and off the air conditioner. 

Zhang and Baillieul in \cite{zhang2012packetized} formalized the concept of \textit{packetized direct load control}, also looking at air conditioner.
According to the solution, individual consumers  can choose their own set points while the operator has the rights to determine the comfort range around such a point.
Based on a thermodynamic model related to the appliance duty-cycle, three theorems were proposed.
The first shows the average room temperature converges to the average set point.
The second indicates a choice for the comfort band.
The third connects the two other theorems, stating that the proposed packetized management is able to achieve the comfort temperature zone with less consumption oscillation, reducing consumption peaks.

The authors extended this initial concept in \cite{zhang2013novel} so that the desired appliance shall now request to consume or withdrawn energy packets according to its own need. 
The process of energy request and withdrawal is designed as queuing system, which has multiple servers and probabilistic returns. 
For the mean waiting time, an analytic expression was derived as function of packet length. 
They showed that a short packet duration with packet switching scheme leads to a mean waiting time smaller than the system without, while the total waiting time for completing the service remains the same. 
Additionally, the packet switching approach provides a fair energy distribution due to its probabilistic nature.

Rezaei and Frolik described in \cite{rezaei2014packetized} a decentralized packetized approach to manage electric vehicles' charging: 
each vehicle needs to request for charging its battery, which can be either approved by a limited time period or rejected. 
This method is adopted from the concept of bandwidth sharing in communication network (e.g. randomized medium access control) and helps to serve all users, respecting the network constraints. 
The proposed packetization was shown to lead to an efficient resource usage, also reducing the travel costs.

Almassalkhi et al. presented in \cite{almassalkhi2017packetized} a real time approach to manage thermostatically controlled loads  that fulfills the grid needs and requires no specific knowledge about the loads' state. 
This proposed scheme is known as packetized energy management, which originates a start-up company  led by the authors and called \textit{Packetized Energy} and a related invention report \cite{frolik2018systems}.
A packetized probabilistic approach has been developed for energy delivery, where each load asynchronously and randomly requests a given server to use an energy packet.
Total population is analyzed so that if there is excess load then the request will be rejected.
The anonymous load requests enable proposed management system to check and analyze an aggregate reference signal.
A case study of thousand simulated water heaters have shown that this process can track the reference signal without noticeably degrading the quality of service (QoS).
The authors recently published a book chapter \cite{almassalkhi2018asynchronous} didactically exposing their achieved results. 

\subsection{Constructing the Energy Internet}

\begin{table}[]
    \caption{Contributions related to the proposed Energy Internet}
    \centering
    \begin{tabular}{|l|l|}
    \hline 
        \textbf{Main topic}           & \textbf{References }        \\\hline\hline 
        Smart Grid development   &   \cite{wang2017distributed,wang2017big,huang2011future,su2013proposing,GaoTIE2018,wang2018survey}        \\\hline
        Packetized energy analysis        &   \cite{saitoh1996new,tsoukalas2008smart,gelenbe2012energy,gelenbe2016energy,gelenbe1994g}        \\\hline
        Physical energy packets     & \cite{abe2011digital,Ma2018optimal,Ma2018elastic,takahashi2015router,rodriguez2015experimental}  \\\hline
        Cyber-physical energy packets        &  \cite{zhang2012packetized,almassalkhi2017packetized,almassalkhi2018asynchronous,zhang2013novel,rezaei2014packetized,frolik2018systems} \\
        \hline
    \end{tabular}
    \label{tab:table}
\end{table}

Although the idea of managing energy systems via packets has been studied and tested in the research presented in the previous subsection and compiled in Table \ref{tab:table}, the focuses differ by looking at macro-level architectural proposals using DC physical packets as in the digital grid \cite{abe2011digital} or on specific micro-level solutions \cite{zhang2012packetized,almassalkhi2018asynchronous}.
We claim that the links between the macro- and micro-levels are still missing and an open field for research.
The recent works by Ma et al. \cite{Ma2018elastic,Ma2018optimal} try to fill this gap using the digital grid concept and therefore is mainly based on physical energy packets.

{Our goal here is different: we would like to build a large-scale cyber-physical system to manage the energy consumption of micro-grids in line with \cite{zhang2012packetized,almassalkhi2018asynchronous}.
To scale up this solution to become dominant in the distribution grid level, micro-grids should be also interconnected, similar to physical packet solution like Abe's and Ma's approaches.
At this point, however, we still need to characterize how the cyber-physical packetized solution is to scale-up to build the Energy Internet.
To reach this goal, we first need to precisely determine what is the energy system and the specific role of the packetized energy within it.}

We assume only the electricity power grid network as illustrated by Fig. \ref{fig:grid} as the energy system to be analyzed here.
In this case, such a system is defined by the operation \textit{electric power interchange}.
In other words, the system only exists (i.e. is considered to be ``alive'') if electricity power flows exist. 
We classify three classes of necessary conditions to keep the system alive \cite{nardelli2017multi}: 
\begin{itemize}
\item \textbf{(1) Condition of production/consumption:} Electric energy needs to be produced (e.g. by power plants or distributed generation) and consumed (e.g. by households, industries etc.). The energy produced need to reach where it is demanded.
\item \textbf{(2) Condition of reproduction/operation:} The generated energy needs to flow at all time to reach where it is demanded. For example, power lines need to be maintained, operational decisions need to ensure power quality, distribution decisions, regulated via market or planing, need to be done etc. In this case, the system ``reproduces'' the state that allows for electricity interchanges.
\item \textbf{(3) External dependencies:} How the system relates to the ``external world'' so that the conditions of production and reproduction can be sustained. For example, investments in electricity production and power lines, training of specialized personal, raw material for power plants, regulatory laws, weather allowing for wind or hydro generation and many other (almost infinite) aspects.
\end{itemize}

\begin{figure*}[!t]
	\centering
	\includegraphics[width=2\columnwidth]{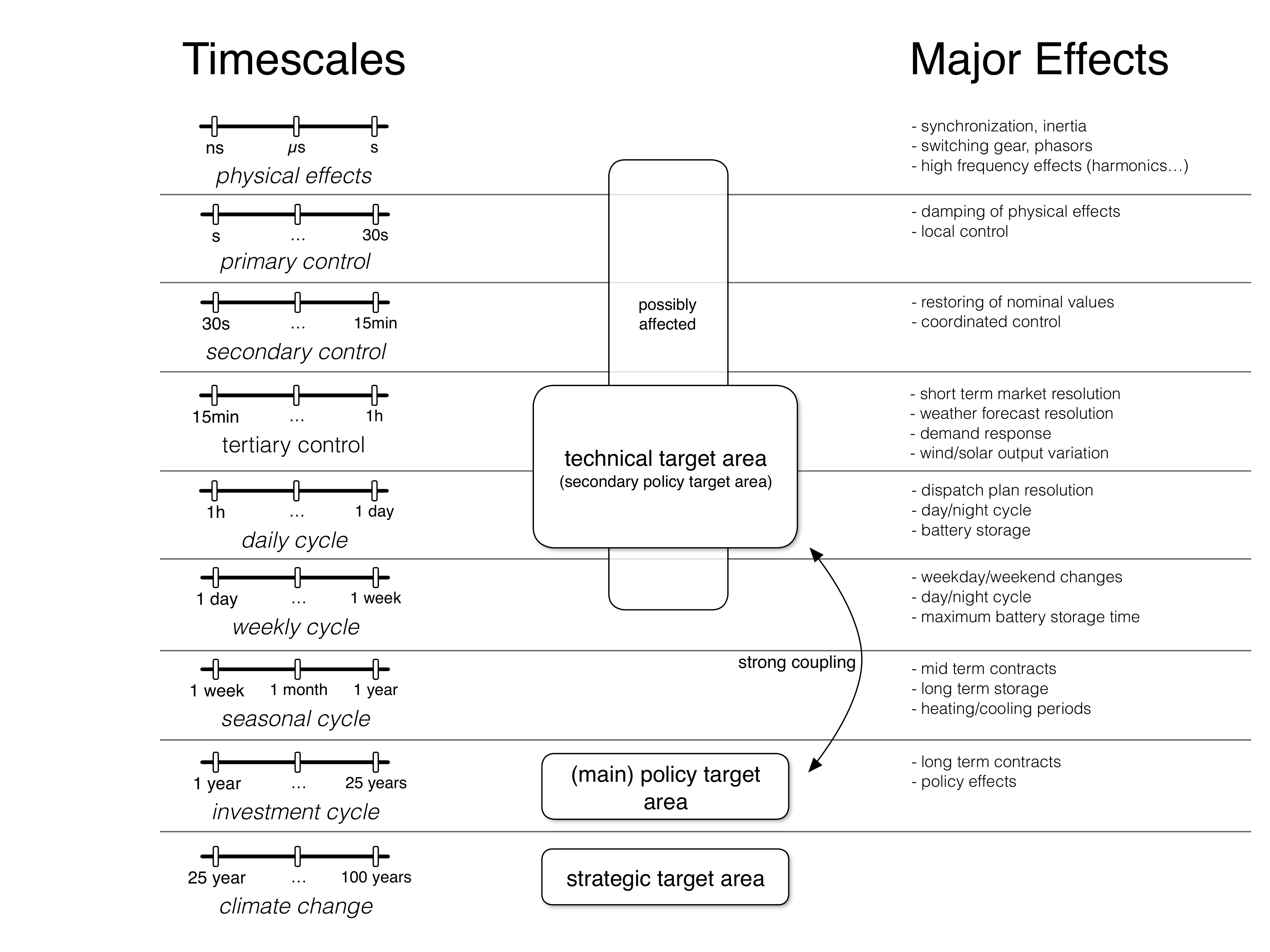}
	\caption{\label{fig:time-scales}A list of different time-scales and related phenomena in electricity power grid analyzed as a system.}
    \vspace{-2ex}
\end{figure*}

The system is analyzed in different time-scales, each of them related to different phenomena.
Fig. \ref{fig:time-scales} lists some examples of it.
Any intervention in the system needs to take the different time-scales into account.
For example, a massive deployment of solar generation implies changes in the frequency control since its way of converting energy is not mechanical.
{The proposed Energy Internet key aspects are related to: 
\begin{itemize}
    \item Micro-grid protection (physical effects, and primary/ secondary control) that will be enabled by the ultra-reliable low-latency regime of machine-type communications;
    \item Packetized energy management algorithm that is the core technical target area, including the required massive connectivity (tertiary control, and daily and weekly cycle);
    \item Different periods of the year as dark Winters with high heating consumption, or bright Summers with low heating consumption (seasonal cycle);
    \item Longer-term investments in micro-grid technologies and distributed generation, considering the possibility of almost zero-marginal cost energy production (e.g. solar and wind) and high storage capacities (investment cycles);
    \item Interventions towards a fossil-fuel free energy system (climate change).
\end{itemize}
}

Next, we will discuss why the Energy Internet is now technologically, socially and economically feasible while exposing the  challenges to guarantee the fulfillment of conditions (1), (2) and (3) so that interchange of electricity can be sustained when large part of the system (distribution network grid in Fig. \ref{fig:grid}) is managed via cyber-physical energy packets.
We will first present the technological developments related to the grid (macro-level) and the specific appliances (micro-level), and how the micro-level shall be organized through virtualiziation via energy packets.
We will then discuss how the recent advances in machine-type communications, data processing and wireless communications in general (IoT networks, 5G and beyond) can fulfill the diverse requirements imposed by the Energy Internet.
Possible new organizing elements that can provide the external support in, e.g. system organizational aspects, as well as in investments and social acceptance, will be also presented.
Finally, we will provide three ongoing research and development activities that are open to incorporate the proposed Energy Internet concept.

\section{Physical grid}
\label{sec:Applications}
%
\subsection{Large-scale power grid}
Power grids have been traditionally consisting of primary power plants in one end, transmission and distribution grids in between and consumer loads in the other end; the power flow has been downwards from the primary power plants \cite{nardelli2014models}. 
In most developed countries, primary loads by industries form the largest part of the total electricity consumption, while the other main part is due to households and commercial buildings.

The situation has lately changed as the distributed energy resources (DER), for instance photo-voltaic (PV) systems, and energy storages (ES) connected to grid have become common, and the number of those have increased both in medium-voltage (MV) and low-voltage (LV) sides in distribution systems. 
Fig. \ref{fig:grid} evinces these changes, which create more interdependences, making power flow control more complex.
On the other hand, such elements can be also used in managing the power balance between production and consumption making the whole power system more flexible. 

Controllable loads in the consumer side together with energy storages are the components that can be used in system control, only inside the LV grid under the MV/LV transformer and/or in bigger scale as part of the whole system power management.
Thus, the whole system can be considered comprising all components in the grid, divided to subsystems containing for instance segments of MV grid, LV grids under the MV/LV transformer, end-customer households, and the loads inside each.
To control the whole system based on any algorithm, it has to be monitored on-line and data must be gathered via sensors and through information-communication technologies (e.g. \cite{wang2018survey}). 
Due to high amount of data and sources, data needs to be processed locally and distributed, and filtered data must flow through a backhaul network, where the operational decisions and control are made. 

To accomplish this, seamless bi-directional data flows via different information and communication layers (data models, communication mediums etc.) are needed. 
These set certain performance requirements for data communications applied in the system(s). 
Communication requirements are dependent on the applications implemented and integrated to power systems, and depending on the application high reliability, high data rates and throughputs, and low latency are required \cite{Wang2011}. 
While grid protection in the transformers requires ultra-reliability and low latency communication \cite{popovski2018wireless}, their generated traffic is rare; electric vehicle charging is more flexible, and coordination between charging stations may not need so strict requirements of latency and reliability; but it requires massive connectivity with security \cite{liu2018sparse}.
More details of specific devices are given next.

\subsection{Specific devices and micro-grids}

Components and devices include all intelligent electric devices (IEDs) interconnected to every consumer loads and small-scale distributed generation (DG) units: wind turbines, PV/solar panels in low-voltage power grid after the MV/LV transformer.
Thus, each low-voltage grid having distributed generation and energy storage (and loads obviously) can be and is considered here as a micro-grid.
These operate as a subsystems, small parts of the whole power grid system.
Inside each micro-grid, IEDs monitor the power flows; power taken from the MV grid, consumption of every load, power generated by every DG, and power reserved to energy storage. 
The concept of digital grid introduced in \cite{abe2011digital}  and very recently extended in \cite{Ma2018elastic,Ma2018optimal} considers that each micro-grid may work independently (asynchronously) and are connected via DC lines where energy is interchanged via physical packets (refer to Section \ref{subsec:phy}).

To build the Energy Internet as proposed here (cyber-physical energy packets), the data from these devices are first collected and then virtualized  as discrete energy packets to local cloud server, which again is connected via some appropriate communication medium to the backhaul network and core cloud servers.
Thus, specialized communication system are to be implemented on every micro-grid. 
Communication technologies inside the micro-grids can be implemented both by wireless IoT sensor networks or power-line communications (PLC) \cite{Pinomaa2011}. 
The selection of appropriate medium and technology is done case-by-case and solution that provides lowest costs. 
Obviously, more important over communication technology costs is that the technology fulfills the performance requirements. 
For instance, grid protection based on relays and those triggering on communications require high reliability and availability, and low latency for communications. 
In \cite{Kansal2012} latency requirement below 100 ms for load and generation trips, and islanding, has been reported. 
In future, the demand for low latency in protection functions will probably become stricter, for instance 1--10 ms is typical response time for machine-to-machine (M2M) (also known as MTC) applications estimated by the European Telecommunications Standards Institute (ETSI) \cite{Noam2012}.
This requires some prioritizing algorithms integrated on the system control, grid protection being the most time-critical application.

Once the physical effects and the primary/secondary control issues are resolved, we could specifically discuss the management of loads and their impact in the grid (tertiary control and daily/weekly cycles).
For example, different households' loads have different priorities that need to be taken into account.
Any management algorithm shall allocate dispatchable loads following a set of rules while guarantee a predetermined quality of service.
In a given household, which load is to be prioritized: the electric cooker or electric car charger? 
Or in a micro-grid community, the dish washer of person \textit{A} or the electric heating of \textit{B}?
The answers depend on both the appliances themselves (their programs, consumption profiles etc.) and their priorities and the electricity cost information (defined by the persons involved, or by some central operator, and/or by the electricity market structure).

\begin{figure*}[!t]
	\centering
	\includegraphics[width=2\columnwidth]{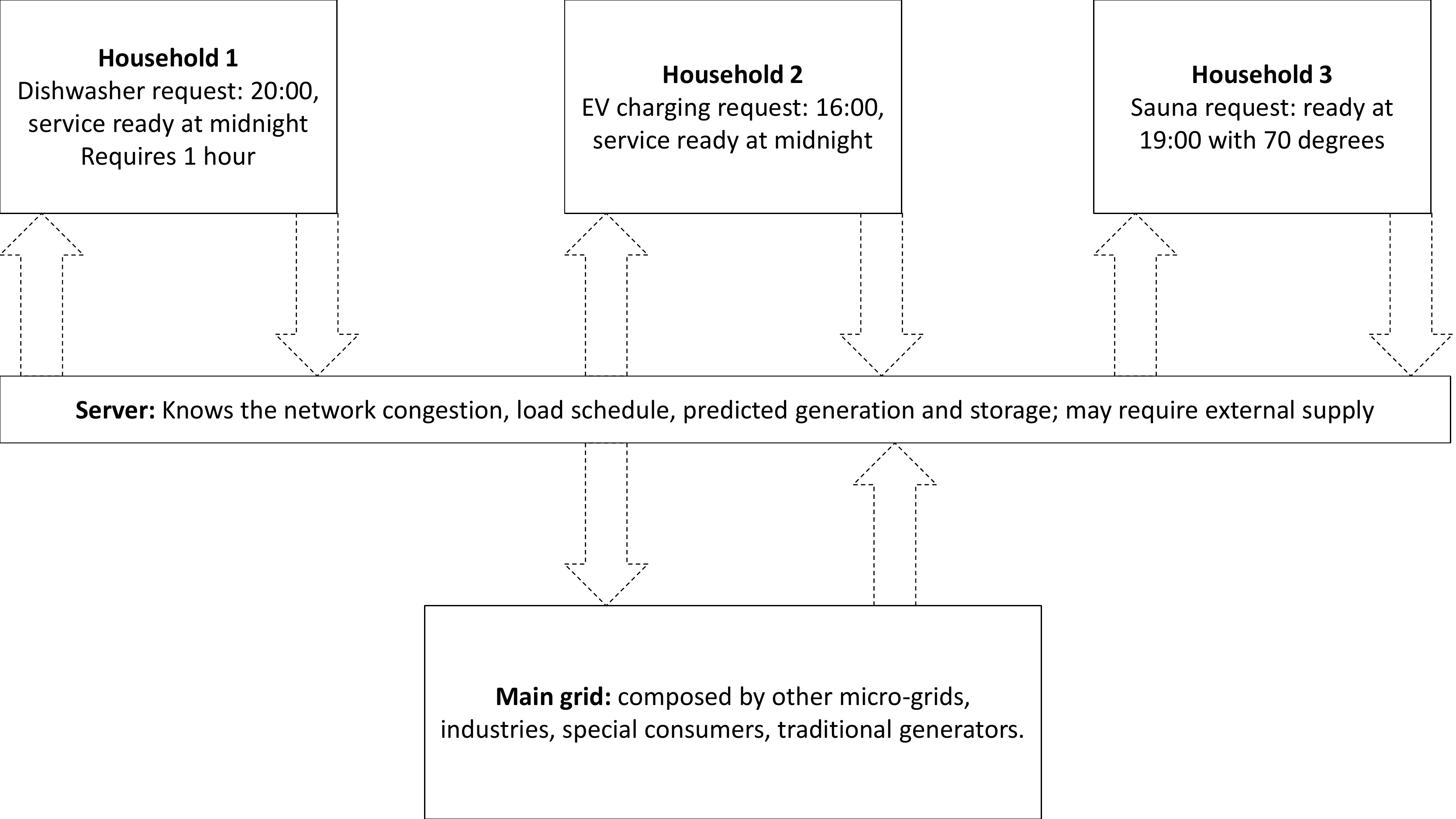}
	\caption{\label{fig:system3}Example of the proposed Energy Internet. There are three households requesting energy packets to be supplied. Household 1 is related to a direct request to use a dishwasher whose program has a predefined energy packet need during a 1-hour period, and the service must be completed by midnight; the request can be only accepted or rejected. Household 2 requests a EV charge starting when the car arrives (e.g. 16:00) and that shall be finished at midnight; the packets can be distributed during this period as far as the final goal is reached. Household 3 requires that a sauna is heated at 70 degrees at 19:00; the way to do this is open (many packets in a short period before, or slow heating through a longer period) and therefore the load is flexible in the period before. The server needs to process these requests based on many factors like the network congestion, predicted generation and storage level. The decision shall be based on the load priorities (e.g. dishwasher has higher priority over the car). The server may also request energy from sources external to the micro-grid so the requests may be satisfied. In case of rejection, the load may request the service once again.}
	%
\end{figure*}

Regardless of the specific arrangement, an effective socio-technical solution to them requires a management system and a communication network capable of coordinating these loads to solve the problem of under-supply or congestion \cite{long2018peer}. 
In this case, the Energy Internet provides the bridge that coordinates in a distributed and fair fashion the micro-decisions so that the desired macro behavior is achieved that must include renewed governance schemes (local market, or a shared pool of energy resources) \cite{van2015power}.
Besides, the digital grid solution may also be used to connect different micro-grids as in Ma et al. \cite{Ma2018elastic}.

\subsection{Grid organization}
To organize and control such complex power grid system in its different time-scales containing all the main power plants and the smallest loads, decentralized and distributed power management system for discretized energy packets transfer is required.
The main idea is to classify different classes of primary users: industrial (e.g. energy intensive), special (e.g. hospitals) and flexible (e.g. households and commercial).
Each specific case shall be related to a specific supply (physically or virtually). 
Let us focus here on the flexible users, specially households.
The supply shall come from distributed sources like PVs, small-scale wind turbines and storages (the energy inventory).
The households are grouped into physical or virtual micro-grids, where the energy resources are shared following a defined set of rules (load priority, demand-response, payment, etc).
The goal might be from minimizing the household costs to maximizing self-sufficiency based on the micro-grid energy inventory; the management algorithm shall be designed to achieve the predefined goal for that specific group of users.

\begin{figure*}[!t]
	\centering
	\includegraphics[width=2\columnwidth]{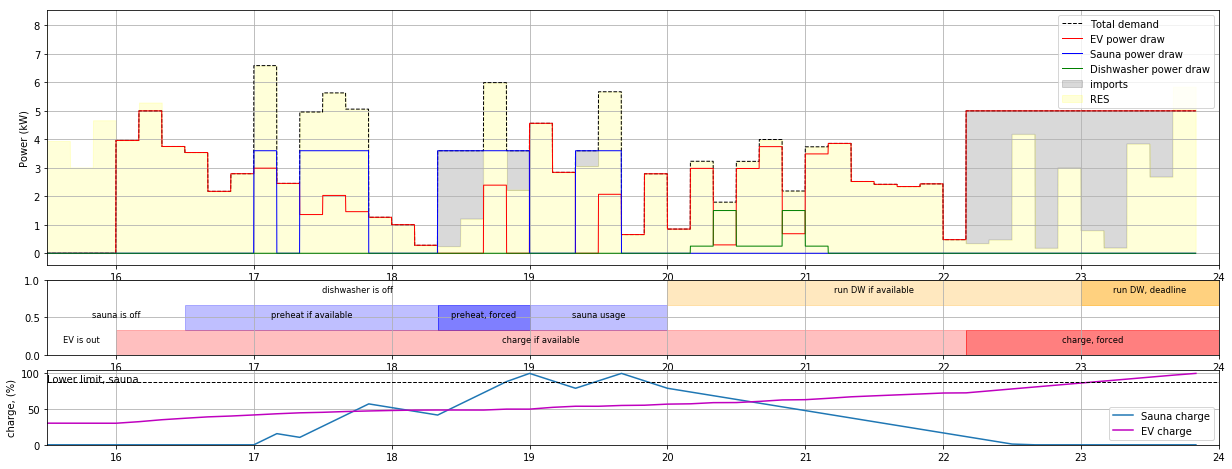}
	\caption{\label{fig:tome}Example of a possible outcome of the case illustrated in Fig. \ref{fig:system3}. The server determines the energy packet allocation of each appliance (i.e. the server indicates which device can access the electric network to consume at most a given amount of energy, pre-determined by the discretized packet size.}
	%
\end{figure*}

Fig. \ref{fig:system3} illustrates this concept, using three households as part of the same micro-grid and an energy server that accept or not their requests.
The Energy Internet would allow this system to work in a distributed and fair manner, as indicated by \cite{almassalkhi2017packetized}.
Before being switched on, IEDs send a requirement of load-usage schedule for a given period. 
This requirement might be served or not, depending on the network conditions, the load priority and the specific household requiring.
As in the internet, the proposed solution shall be based on fairness via randomization and net-neutrality in relation to the users, but not to the loads as far as some loads must have priorities over others.
What determines the load priority relates to the group decision, or the service provider.
In situations that the micro-grid cannot be self-sustained, namely over- or under-supply, the connection to the main grid needs to be available. 

Fig. \ref{fig:tome} exemplifies one possible outcome from the Energy Internet management based on requests and a server.
The three loads are modeled so that: (1) Sauna is a thermal load (dispatchable) with fixed power of 3600 watts); (2) EV arrives at 16:00 with a random initial state and can be charged any value between 0 and 5000 watts, with a battery with capacity of 30 kWh; (3) Dishwasher fixed cycle with one hour length, dispatchable from 20:00.
It was also simulated a random sequence to model the renewable sources (solar plus wind).
In the top plot, the individual loads are the solid lines while the black dashed lines are the aggregate demand.
Yellow shadowing indicates local renewable energy, while the gray shadowing indicates the imported energy needed in order to supply the demand.

The server accepted the three loads and provided the pre-allocation given in the middle plot by devices: (1) dishwasher may run from 20:00 depending on the energy availability and network capacity constraint, but it must run at 23:00 latest; (2) sauna can be preheated from 16:30 onward, but if it has not yet reached the sufficient temperature level at 18:20 it goes through a ``forced'' heating regime so it can be used between 19:00 and 20:00; (3) EV can be charged from 16:00 onward with a forced charge from 22:10 (so the charge can be completed at mid-night.
Note that the loads have different kind of flexibility.
The EV needs to be fully charged to be in used at a given time.
The dishwasher has flexibility when it starts, but once it is running flexibility ends.
Sauna is a thermal load that needs to be at a given temperature at a specific time so that, for example, it could be heated much above the needed temperature two hours before the use, or slowly heated (consuming less electric power) for a longer period (four hours), or normally heated to the specific temperature twenty minutes before the use.
In any case, all loads (once accepted) have a deadline period when the charge must take place, leading to a ``forced use'' regime.
The server must coordinate these loads based on planning and operational balancing, as indicated by the bottom plot about charging level from the EV and sauna.

As in the liberalized markets, this shall happen in different time-scales similar to the day-ahead and balancing markets.
It is interesting to note that, in European electricity markets, there exist the so-called Exclusive Group (e.g. \cite{kuhnlenz2018implementing}) where daily profiles are bid, not independent hour-by-hour bids.
In this class of bidding, the load profile selected by the market matching algorithm shall be realized.
In this sense, Exclusive Group requires flexibility and the proposed Energy Internet would be a suitable.
Therefore, the proposed solution can be already implemented within the existing market structure.
In critical situations, the system shall enter in emergency mode guaranteeing the basic energy needs.
All in all, the physical grid -- not the virtualized one -- is the determinant in the last instance.

\section{Machine-type communications}
\label{sec:MTC}
Ubiquitous wireless connectivity is a key enabler of the Energy Internet, and is the ultimate goal of the 5th generation of wireless networks (5G). 
Differently from previous generations, 5G inherently encompass MTC, which  involves a wide range of heterogeneous applications with different requirements \cite{Nokia2016}.
For example, in modern power grids applications, they range from simple daily electricity metering to advanced real-time frequency control. 
To cover such diverse requirements, MTC needs to work in distinct modes related to the specific application under consideration \cite{5GEnergy}.

To address such wide range of applications MTC is split into two poles: massive MTC (mMTC) and ultra-reliable low-latency communications (URLLC) \cite{NokiacMTC2016}. 
As the names suggest, mMTC incorporates applications with a very large number of connected devices (connectivity goal) \cite{liu2018sparse} while URLLC focuses on mission critical (high reliability and low latency goal) communication \cite{popovski2018wireless}. 
These two new modes enable MTC networks to fulfill heterogeneous requirements from massive connectivity up to ultra-reliability and low latency, which cannot be attended by current commercial technologies nor by the conventional techniques used in  broadband wireless communications. 
In this context, MTC is a key enabler of the Energy Internet by autonomous exchange of information; depending on the application, it can use either a mMTC technology so that IEDs can be almost anywhere or a URLLC technology to achieve strict latency and reliability imposed by the energy grid in order to keep the electricity interchanges happen within the quality constraints \cite{Kuzlu2014,5GEnergy}.

\begin{figure*}[!t]
	\centering
	\includegraphics[width=2\columnwidth]{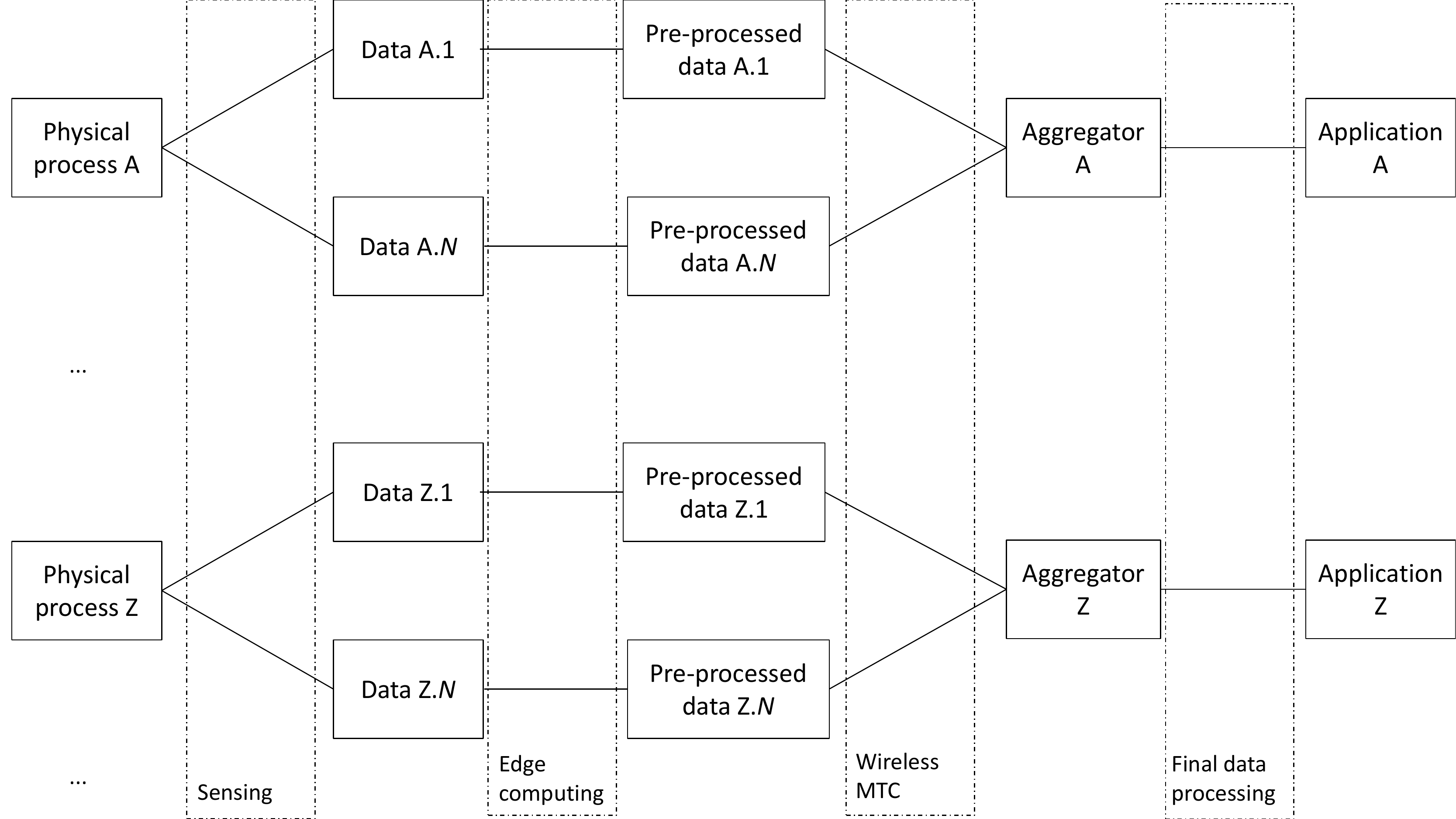}
	\caption{\label{fig:system4}Illustration of a machine-type communication framework for the Energy Internet. There are many physical processes related to the system happening at the same time. Sensors are deployed to acquire raw data. This data is pre-processed in the edge following a given strategy (e.g. time-based or event-based sampling). The pre-processed data is aggregated/fused and then transmitted via wireless MTC. At this stage, depending on the application requirements, the transmissions must be related to mMTC (e.g. metering) or URLLC (e.g. fault detection). The final step is processing of the received data to be used by the end-application, which may be the Energy Internet server or an automatic voltage/frequency control feedback loop.}
	\vspace{-1ex}
\end{figure*}

A promising way to deal with the massive connection problem comes from the concept of data aggregation. 
Traffic from MTC devices -- which may or may not be pre-processed -- is first transmitted to a data aggregator that collects and processes the received data. Depending on the application, the aggregator can relay the processed data optimizing signalization to the core/backhaul network \cite{8120238MMTC_LTE}, use the data as part of a feedback control loop, or be used to monitor some metric (e.g. average energy consumption). 
Both congestion and the power consumption of the devices are expected to decrease, while (spatial) spectral efficiency can be improved. 
When aggregating several devices, the density of the aggregators (even though considerably smaller than the devices’ density, but still large) and the interference coming from the devices sharing the same resource could be significant and is a challenging problem \cite{Lopez2017}.
The performance of the communication network can also improve by incorporating some intelligence to the aggregator as, for example, resource scheduling \cite{Guo2017}. The number of served devices, on the other hand, may create the so-called scalability problem where there exist a limited number of resources available at the aggregator. This reduces both energy and spectral efficiency. To combat such losses, non-orthogonal multiple access strategy is, for instance, an attractive solution; this scheme supports multiple users that can be multiplexed at different power levels but at the same time/frequency/code with a satisfactory performance \cite{Lopez2017,Shirvanimoghaddam2017b}. 

Besides, traditional communication systems are often designed and optimized using the notion of channel capacity that assumes infinite long block lengths, which is a reasonable benchmark for practical systems with rely on large averages to project the data rates \cite{Durisi2016a}. However, MTC (either mMTC or URRLC) traffic is often characterized by short blocks. As a result, the same assumptions in terms of channel capacity cannot be directly applied to short block length messages as pointed out in \cite{Durisi2016a}, which surveys the recent advances in this area. In this context, new information theoretic results have been presented related to the performance of short block length communication systems showing that achievable rate depends not only on the channel quality of the communication link, but it is also a function of the actual block length and error probability tolerable at the receiver. In light of these new results, the underlying channel models are being revisited for IoT specific environments, including the traffic characterization and energy consumption constraints.  

Effective capacity theory appears in this context as a cross-layer metric that offers guarantees on statistical QoS provisioning subject to latency constraints. Effective capacity provides the maximum constant arrival rates at the link layer that can be supported with a time-varying wireless transmission rates while guaranteeing certain delay/latency (QoS) constraints. Delay-bounded QoS is an important metric for delay-bounded MTC networks due to the diverse traffic generated by different applications with their heterogeneous requirements \cite{Ozmen2016}. For instance, as commented above protection signals must be dealt within few milliseconds, data collection from smart meters may have tens of milliseconds of latency, and, in the other extreme, average consumption monitoring may accepted latency of minutes. Moreover, the characteristics of the traffic generated by the those different applications and their constraints onto underlaying network performance. 

Figure \ref{fig:system4} illustrates a general machine-type communication framework, where different physical processes running in parallel require different performance from the communication system depending on the final application.
In the case of the Energy Internet, the physical processes are virtualized so as to map energy consumption, generation and storage as discretized packets.
Processes related to, for example, grid protection are virtualized and processed to guarantee that the operation of the electric grid is under their strict constraints -- i.e. high reliability and low latency.
Overall, many advances have been made over the recent years, and MTC is seen as key enabler for future wireless networks, 5G and IoT, and such accomplishments are a fundamental enabler of the Energy Internet with its diverse requirements related to delay, reliability, coverage and connectivity. 

\section{Other issues}
\label{sec:other-issues}
Managing energy systems as packets not only brings  technological challenges related to communications and the grid itself but also opens questions on how the Energy Internet could be sustained in other dimensions.
This would go from social acceptance of the innovative solution to new business models since it presupposes active management strategies supported by technological advances for monitoring and controlling the distribution network over a reliable communication infrastructure.
For example, nowadays there is a wide range of emerging arrangements (e.g. \cite{giordano2012business,gregoratti2015distributed,huang2019embedding,van2015power,long2018peer} and references therein) that support local energy trading, flexible management and localized control of electricity networks.
Their actual suitability to the proposed solution is not straightforward, but would be seen as start points to be further investigated and eventually deployed.

In \cite{giordano2012business}, the authors indicated that the recently developed grid technologies, the regulation bodies (i.e. laws) and business models are not synchronized, creating obstacles to a systemic modernization of the grid \cite{nardelli2014models}.
Despite of these impairments, some directives and standardization bodies like 5G-PPP \cite{5GEnergy} understand that modernizing solutions will be built upon a more cooperative economy \cite{van2015power}, opening the door to new  solutions. 
In this context, three interesting concepts are worth mentioning as potential (social) elements of the Energy Internet.
They are the following:
\begin{itemize}
\item \textbf{Micro-operator} (e.g. \cite{pouttu2017p2p}) can be considered as an entity that offers mobile connectivity combined/locked with specific, local service, which is spatially confined to the defined area of operation, such as micro-grids. Micro-operator model would be used as the way to manage the communication network to sustain the Energy Internet. It can be private, public or community owned.
\vspace{1ex}
\item \textbf{P2P aggregator} (e.g. \cite{pouttu2017p2p}) is built to become a new type of supplier that brings together residential, commercial and industrial demand and facilitate the integration of demand-side flexibility in the existing electricity market. The P2P aggregator of a given group serves as both: manager of flows within the groups and the link between between the group and the external grid. It may operate the P2P trading platform, which creates a marketplace providing information and matching buyers and sellers, allowing the self-regulation of the P2P trading. This concept was studied in the H2020 P2P-SmarTest. 
\vspace{1ex}
\item \textbf{Energy community/local cooperative:} (e.g. \cite{van2015power,huang2019embedding})  is a community-oriented entity that finances projects on renewable installations. There can be tens of thousands of members (e.g. Ecopower cooperative in NobelGrid project has 50.000 members). The electricity produced with these installations is supplied to the cooperative members. This model can manage renewables and serve members, while operating in an open, competitive market environment. It places \textit{prosumers} (consumers who are also small-scale producers) in the center of local energy market design. Small to medium-scale distributed storage systems could be used to exploit their flexibility. This allows for cooperatives based on physical location (e.g. micro-grids) or virtual associations (similar to Virtual Power Plants). There are many projects about community energy as, for example in Europe, H2020 NobleGrid, WiseGRID and Empower.
\end{itemize}

These elements indicate the diversity of possible participants in the future energy system.
For example, the P2P aggregator may act as the ``energy server'' that decides if energy packet requests will be served or not; it shall consider not only the availability of generation/storage to serve the request, but also the load that request (load priorities, e.g. cooking has higher priority than electric vehicle charging) and the network capacity, as well as the source of supply.
If we are considering a micro-grid operating as a local cooperative that tries to be off-grid as much as possible, then the optimization shall manage the ``energy inventory'' based on such a goal.
The communication network operated by the micro-operator needs then to fulfill the grid requirements so that the P2P aggregator could properly manage the local energy cooperative. 
So it would fulfill the communication requirements needed to manage the interchange of energy packets locally, also allowing for external communication when the micro-grid needs to be connected to the main grid (due to over- or under-supply).

One advantage of local cooperatives operating as a micro-grid is that users (\textit{prosumers}) tend to be engaged with the management process, more than usual relations between consumer and utility, or consumer and service providers \cite{van2015power}.
Due to the nature of cooperatives, the decision-making and directives of the micro-grid would be decided collectively.
For example, the community would decide (bounded by technical and economical limitations) about the loads' priorities, investments in generation, acquisition of storage, and how to deal with surpluses or deficits of energy.
Due to the democratic process of decision-making, the social acceptance of the energy management system based on packets tends to increase (e.g. \cite{van2015power}).

Interestingly, issues related to social acceptance in energy systems is the topic of the first issue of IEEE Power \& Energy Magazine in 2018, including two articles \cite{steg2018drives,perlaviciute2018heart} that also identify the importance of active participation in decision-making to engage participation of users in, for example, demand-response programs.
They argue demand-response policies that require changes in energy consumption (e.g. turning off air conditioner, or delay washing machines, or randomized approach to charge electric vehicles at nights) would be more likely accepted when the users feel they actively participate in the system when compared to more top-down, price-based, approaches.

\section{Ongoing and future activities}
\label{sec:future}

This paper provides our perspective of how to manage the distribution level of electricity power grids as cyber-physical systems based on discretized energy packets.
In addition to the main results already discussed in the previous sections (mainly \cite{almassalkhi2018asynchronous,abe2011digital,Ma2018elastic,Ma2018optimal}), we would like to cite three ongoing activities from the authors that set the basis of the proposed concept.
They are: BCDC Energia, zero-energy log house, and Fusion Grid.
We will provide a brief summary and their relation with the Energy Internet in the following.

\subsection{BCDC Energia}
  
The consortium named \textit{Cloud Computing as an Enabler of Large Scale Variable Distributed Energy Solutions -- BCDC Energia}\footnote{\url{http://www.bcdcenergia.fi/en/project/}} focuses on identifying different solutions to guarantee the balance between supply and demand using a high quality weather forecast via state-of-the-art information and communications technologies.
One line of research is the development of an energy-weather forecast by the Finnish Meteorological Institute as a tool to predict the electricity production from solar and wind with their most recent meteorological model.
Other research line is the analysis of machine-type communications and IoT networks in the context of energy system as in \cite{ramezanipour2018increasing,ramezanipour2018finite,alves2018secure,lopez2018aggregation,tome2018long,tome2018event}.
In more recent work \cite{tome2018storage}, Tomé et al. gave the first step to combine them by focusing on how weather forecast and  statistical analysis of electricity demand could be the basis of energy storage management in households and micro-grids.

In more systemic level, the impact of demand-side management was studied in \cite{kuhnlenz2018demand,kuhnlenz2018implementing,kuhnlenz2016dynamics,nardelli2017smart,kuhnlenz2017agent} following the lines presented in \cite{nardelli2014models}.
These results evince the existence of systemic emergent phenomena that are undesirable when operating power grids, mainly due to a combination between lack of coordination caused by poorly designed information signals.
The authors provide possible solutions to the issue, including direct signals for micro-grids \cite{kuhnlenz2018demand} and exclusive block bids in wholesale market \cite{kuhnlenz2018implementing}.

The results from BCDC Energia are, however, mainly based on empirical or theoretical research supported by already available data with neither explicit testbed to assess the concept in real-case scenarios nor system-level simulations to prove its scalability.
The packetized energy has already showed its potential as a specialized solution (refer to Section \ref{sec:EnergyIntenet}) that can be used to combine the different research direction opened by BCDC Energia including small-scale testbeds (households and micro-grids) and simulations/emulations for scaling it up.
Those aspects are covered in, for example, the following two research activities.

\subsection{Zero-energy log house}

The \textit{nollaenergiahirsitalo} is the first organized zero-energy log house in Finland\footnote{More information at \url{http://www.nollaenergiahirsitalo.fi/} (in Finnish).}.
The construction was led by one of the co-authors (Antti Kosonen) so that the aggregate energy consumption and demand is zero.
To do so in the challenging Finnish conditions with cold and dark winters, not only an energy efficient building is needed but also an intelligent management algorithm, as described next. 
The buildings in the site are positioned to guarantee shadowless of the roofs in order to maximize solar photo-voltaic (PV) production.
The house has south and the garage east-west solar PV installation, including total of 21 kWp modules and 16 kW of inverter power with four separate maximum power point tracking. 
The main heating system is based on a ground source heat pump with vertical drill hole that enables also pre-cooling of the incoming air. 
The heat is transported through the water that circulates in the floors. 
The heat can be storage to a 750-liter, two-layer water boiler, in which the upper layer is mentioned for domestic hot water and the lower layer for buffering (storage) of the heating. 
In addition, the lower layer is applied as a pre-heater of the domestic hot water. 
The heat pump is controlled with a Rasperry PI system that utilizes solar PV production forecast based on energy-weather forecast of Finnish Meteorological Institute (see BCDC Energia) and wholesale market electricity price (given by NordPool).

\begin{figure*}[!t]
	\centering
	\includegraphics[width=2\columnwidth]{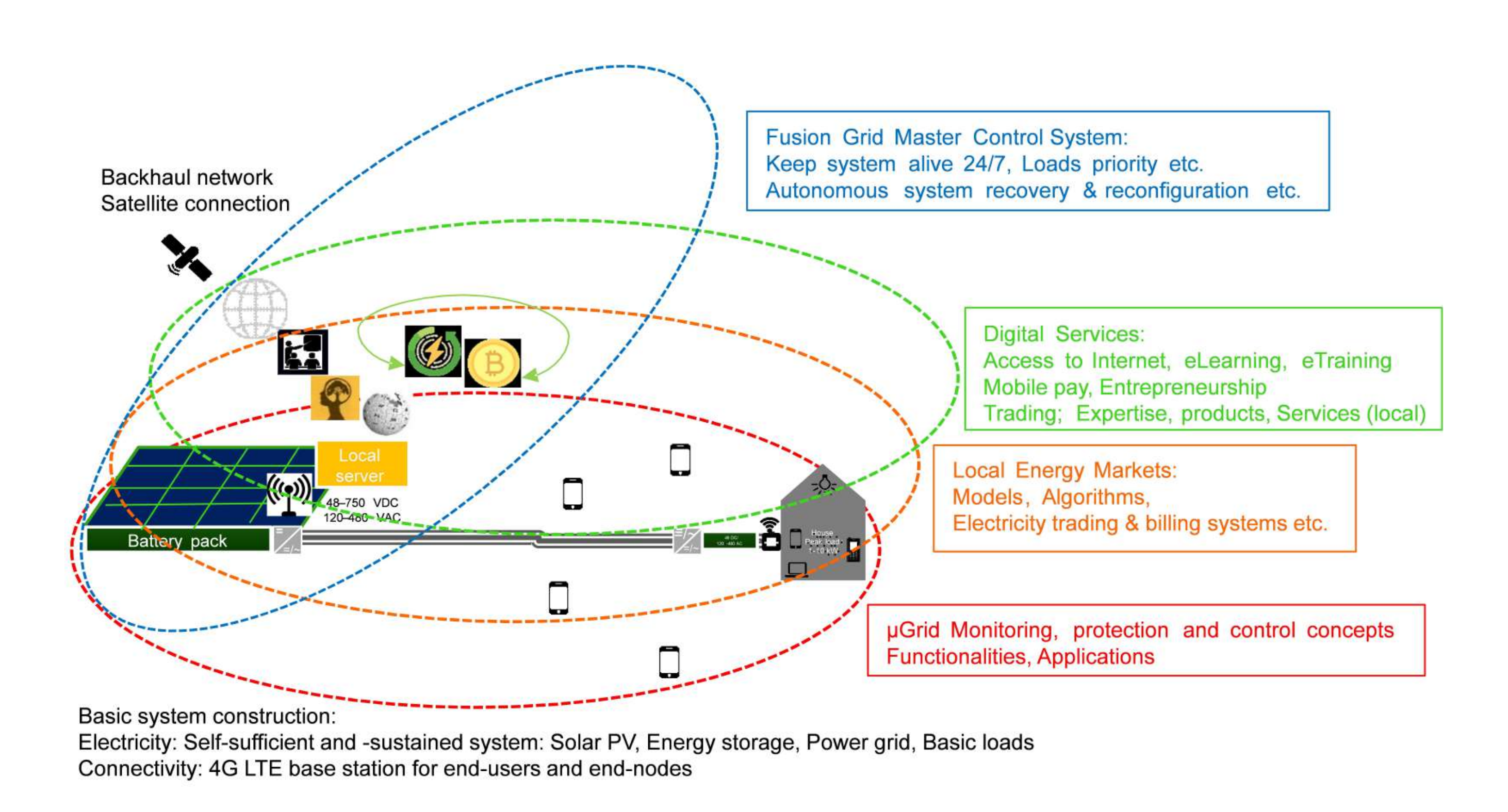}
	\caption{\label{fig:FG} Fusion Grid schematic composed by electricity micro-grid and wireless connectivity, including possible digital services and local markets.}
\end{figure*}

The intelligence introduced by the control system is able to perform well in the present situation.
However, we have shown that this solution cannot be sustained when many more houses use the same kind of intelligent algorithm (i.e. it is not a scalable solution at the existing energy system structure).
Coordination is needed for either case: micro-grids \cite{kuhnlenz2018demand,nardelli2017smart} or liberalized markets
\cite{kuhnlenz2018implementing,kuhnlenz2017agent}.
The packetized management with requests and a server will then serve to coordinate such intelligent agents using distributed solutions like the ones proposed in \cite{pilgerstorfer2017self,pournaras2017self,nambi2016temporal} that can solve collective decision-making processes with combinatorial
complexity.
This solution also allows market integration as discussed in \cite{kuhnlenz2018implementing}.

\subsection{Fusion Grid}
Fusion Grid\footnote{More details in \url{https://bit.ly/2OSFhMn}.} is a project supported by Business Finland BEAM program and led by Lappeenranta University of Technology (LUT). Other project partners are Nokia, GreenEnergy Finland (GEF), University Properties of Finland (SYK), and Aalto Yliopisto. 
The main goal in the project is to develop a cost-effective solution to bring electricity (via off-grid micro grid based on renewable energy sources) and connectivity (LTE Kuha base-stations developed by Nokia) in remote and poor areas where there is no existing network infrastructure \cite{demidov2018fusion}.
This solution involves a light off-grid cell deployment comprising solar panels, battery energy storage, power delivery solution for local people and houses with few controllable loads with wireless connectivity and a Kuha base-station.
The basic construction of the Fusion Grid platform consists of different levels with functionalities as illustrated in Fig. \ref{fig:FG}.

The off-grid power network is designed to be self-healing and self-configured when there is fault in the power system, or when/if the grid is extended. 
All grid-related control and critical communications are wireless and coordinated via Kuha base-station. 
This requires intelligent devices to be installed not just to next to the Kuha base station (i.e. the grid master control unit) but also to the each control point (i.e. the customer power grid interfaces). 
To be a cost-effective off-grid power system implementation, customers may also have their own small-scale PVs and/or energy storages making the power system decentralized. 
Accordingly, there might be a need and interest to exchange the energy with the rest of the power grid.
This involves and requires local off-grid energy markets and models to be designed for such systems. 
These make the power grid also decentralized in relation to the grid control and communications so that and event-based communication algorithms with local data pre-processing algorithms need to be studied. 
Cost efficiency is also the aim in communications via the Kuha base station, which also provides access to internet and local services for customers. 
Thus, the same wireless communications network are used for both customers, and control of the power system.
Two testbeds are expected: one at LUT Green Campus, where the off-grid power system and the control concepts of Fusion Grid are designed and tested; another will be implemented in Namibia, including active feedback from end-users.

As the first test setup is to be built, the project focuses on few loads and a reasonably small number of households so that demand-side coordination is still not an explicit concern.
However, the Fusion Grid is being developed to be modular so it shall be scalable so that it can be extended by interconnecting similar small-sized power cells (``nano-cells'') together, and by this way wider-scale energy distribution infrastructure could be formed.
In this sense, Energy Internet offers a direct solution to coordinate loads via Kuha.
Micro-grid functionalities, such as communication-based grid protection, where latency and reliability demands are very strict are developed and tested on the pilot setup. In addition, autonomous (1) grid recovery from a fault in the power system, and (2) grid reconfiguration when power system  with a new nano-cell is extended will be modeled and emulated by simulation models.
Another possible approach is to emulate different households based on the zero-energy log house data and test its scalability via simulation and emulation in the Fusion Grid platform. 
These would include different possibilities of governance such as self-sufficient energy communities, local micro-grid markets or fully market-integrated virtual micro-grids.

\section{Discussions and final remarks}
\label{sec:final}
This article discusses how to build the Energy Internet supported by the recent technological developments.
By revisiting the relevant literature, we demonstrated the reasons why managing the distribution level of the electricity grid based on cyber-physical energy packets is feasible and desirable.
Our approach considers that users, being consumers and/or producers with storage capabilities, are part of a (physical or virtual) micro-grid and possess a common inventory that needs to be managed by, for example, a P2P aggregator who acts as a server.
The inventory shall have rules that include priority of loads and optimization goals (e.g. being self-sufficient at most times); these can be decided collectively by the micro-grid members or dictated by the system operator.
The proposed Energy Internet can be only sustained by the massive connectivity and the ultra-reliability low-latency guarantees given by machine-type communications that fulfill the strict quality requirements imposed by the physical grid.
In this sense, machine-type communications  become a necessary enabler to build the Energy Internet.
However, technology alone is not sufficient so that different elements like micro-operator and aggregators are then needed.
Together with these elements, social acceptance of the proposed solution is necessary.

Clearly, the vision presented here is ambitious and may take decades to be fully realized.
However, a transition from existing liberalized markets to this approach can be done via virtual micro-grids that treat their energy according to principles described here while participating in the wholesale market (probably together with other micro-grids).
This approach is also aligned with the solutions proposed in \cite{kuhnlenz2018demand,kuhnlenz2018implementing}, where the P2P aggregators would play a bigger role by creating possible local energy interchanges that would lead to the upcoming Energy Internet.
{Overall, the proposed Energy Internet can only emerge if the following open research topics are tackled:
\begin{itemize}
    \item \textbf{Packetized Energy Management:} It is necessary to research the best ways to discretize dispatchable loads to be handled by an energy server, whose management algorithm can solve (large) combinatorial problems in few minutes. Extending the work in the management side done by \cite{almassalkhi2018asynchronous} into the direction of the distributed algorithms in \cite{pournaras2017self} indicates a promising solution.  
    \item \textbf{Micro-grid:} The key open questions in this topic are related to the effects of communications delay on micro-grids, mainly related to protection, and how the interconnection between micro-grids shall be done. The surveys presented in \cite{hosseini2016overview,Kuzlu2014} indicate different research paths. We are particularly interested in the concept of Fusion Grid \cite{demidov2018fusion} presented in Fig. \ref{fig:FG} as a modular solution where the Packetized Energy Management can be tested using wireless communications in a micro-grid environment.
    \item \textbf{Machine-type communications (MTC):} The research in the field is normally application agnostic \cite{liu2018sparse,popovski2018wireless}. It is necessary that wireless communication technologies are tailored to cover the Energy Internet applications. In general, reference \cite{wang2018survey} provides a survey related to the evolution of energy management including ICTs. More focused research in applications of MTC in energy systems are given in \cite{ramezanipour2018finite,tome2018long}.
    \item \textbf{Governance models:} A new system paradigm requires different governance models. The liberalized electricity market composed by big generators and retailers are becoming unsuitable for the new energy system when more and more distributed generation are appearing at the distribution level. New governance models like energy communities \cite{van2015power}, or peer-to-peer local markets \cite{long2018peer}  need to be further studied and deployed to sustain  the proposed Energy Internet.
\end{itemize}}

\bibliographystyle{IEEEtran}
\bibliography{ref}

\begin{thebibliography}{10}
\providecommand{\url}[1]{#1}
\csname url@samestyle\endcsname
\providecommand{\newblock}{\relax}
\providecommand{\bibinfo}[2]{#2}
\providecommand{\BIBentrySTDinterwordspacing}{\spaceskip=0pt\relax}
\providecommand{\BIBentryALTinterwordstretchfactor}{4}
\providecommand{\BIBentryALTinterwordspacing}{\spaceskip=\fontdimen2\font plus
\BIBentryALTinterwordstretchfactor\fontdimen3\font minus
  \fontdimen4\font\relax}
\providecommand{\BIBforeignlanguage}[2]{{%
\expandafter\ifx\csname l@#1\endcsname\relax
\typeout{** WARNING: IEEEtran.bst: No hyphenation pattern has been}%
\typeout{** loaded for the language `#1'. Using the pattern for}%
\typeout{** the default language instead.}%
\else
\language=\csname l@#1\endcsname
\fi
#2}}
\providecommand{\BIBdecl}{\relax}
\BIBdecl

\bibitem{elayoubi20175g}
{S. E. Elayoubi \textit{et al.}}, ``{5G innovations for new business
  opportunities},'' in \emph{Mobile World Congress}.\hskip 1em plus 0.5em minus
  0.4em\relax 5G Infrastructure Association, 2017.

\bibitem{wang2017wireless}
K.~Wang, Y.~Wang, X.~Hu, Y.~Sun, D.-J. Deng, A.~Vinel, and Y.~Zhang, ``Wireless
  big data computing in smart grid,'' \emph{IEEE Wireless Communications},
  vol.~24, no.~2, pp. 58--64, 2017.

\bibitem{wang2017distributed}
K.~Wang, L.~Gu, X.~He, S.~Guo, Y.~Sun, A.~Vinel, and J.~Shen, ``Distributed
  energy management for vehicle-to-grid networks,'' \emph{IEEE Network},
  vol.~31, no.~2, pp. 22--28, 2017.

\bibitem{wang2017big}
K.~Wang, H.~Li, Y.~Feng, and G.~Tian, ``Big data analytics for system stability
  evaluation strategy in the energy internet,'' \emph{IEEE Transactions on
  Industrial Informatics}, vol.~13, no.~4, pp. 1969--1978, 2017.

\bibitem{abe2011digital}
R.~Abe, H.~Taoka, and D.~McQuilkin, ``Digital grid: Communicative electrical
  grids of the future,'' \emph{IEEE Transactions on Smart Grid}, vol.~2, no.~2,
  pp. 399--410, 2011.

\bibitem{Ma2018optimal}
J.~Ma, L.~Song, and Y.~Li, ``Optimal power dispatching for local area
  packetized power network,'' \emph{IEEE Transactions on Smart Grid}, 2018.

\bibitem{Ma2018elastic}
J.~Ma, N.~Zhang, and X.~Shen, ``Elastic energy distribution of local area
  packetized power networks to mitigate distribution level load fluctuation,''
  \emph{IEEE Access}, vol.~6, pp. 8219--8231, 2018.

\bibitem{saitoh1996new}
H.~Saitoh and J.~Toyoda, ``A new electric power network for effective
  transportation of small power of dispersed generation plants,''
  \emph{Electrical engineering in Japan}, vol. 117, no.~1, pp. 19--29, 1996.

\bibitem{zhang2012packetized}
B.~Zhang and J.~Baillieul, ``A packetized direct load control mechanism for
  demand side management,'' in \emph{IEEE 51st Annual Conference on Decision
  and Control (CDC)}, 2012, pp. 3658--3665.

\bibitem{almassalkhi2017packetized}
M.~Almassalkhi, J.~Frolik, and P.~Hines, ``Packetized energy management:
  {A}synchronous and anonymous coordination of thermostatically controlled
  loads,'' in \emph{IEEE American Control Conference (ACC)}.\hskip 1em plus
  0.5em minus 0.4em\relax IEEE, 2017, pp. 1431--1437.

\bibitem{almassalkhi2018asynchronous}
M.~Almassalkhi, L.~D. Espinosa, P.~D. Hines, J.~Frolik, S.~Paudyal, and
  M.~Amini, ``Asynchronous coordination of distributed energy resources with
  packetized energy management,'' in \emph{Energy Markets and Responsive
  Grids}.\hskip 1em plus 0.5em minus 0.4em\relax Springer, 2018, pp. 333--361.

\bibitem{economist}
\BIBentryALTinterwordspacing
``{Building the energy internet}.'' [Online]. Available:
  \url{http://www.economist.com/node/2476988}
\BIBentrySTDinterwordspacing

\bibitem{tsoukalas2008smart}
L.~Tsoukalas and R.~Gao, ``From smart grids to an energy internet: Assumptions,
  architectures and requirements,'' in \emph{Electric Utility Deregulation and
  Restructuring and Power Technologies, 2008. DRPT 2008. Third International
  Conference on}.\hskip 1em plus 0.5em minus 0.4em\relax IEEE, 2008, pp.
  94--98.

\bibitem{huang2011future}
A.~Q. Huang \emph{et~al.}, ``The future renewable electric energy delivery and
  management (freedm) system: {T}he energy internet,'' \emph{Proceedings of the
  IEEE}, vol.~99, no.~1, pp. 133--148, 2011.

\bibitem{su2013proposing}
W.~Su and A.~Q. Huang, ``Proposing a electricity market framework for the
  energy internet,'' in \emph{Power and Energy Society General Meeting (PES),
  2013 IEEE}.\hskip 1em plus 0.5em minus 0.4em\relax IEEE, 2013, pp. 1--5.

\bibitem{GaoTIE2018}
M.~Gao, K.~Wang, and L.~He, ``Probabilistic model checking and scheduling
  implementation of energy router system in energy internet for green cities,''
  \emph{IEEE Transactions on Industrial Informatics}, vol.~PP, no.~99, pp.
  1--1, 2018.

\bibitem{wang2018survey}
K.~Wang, J.~Yu, Y.~Yu, Y.~Qian, D.~Zeng, S.~Guo, Y.~Xiang, and J.~Wu, ``A
  survey on energy internet: Architecture, approach, and emerging
  technologies,'' \emph{IEEE Systems Journal}, vol.~12, no.~3, pp. 2403--2416,
  2018.

\bibitem{takahashi2015router}
R.~Takahashi, K.~Tashiro, and T.~Hikihara, ``Router for power packet
  distribution network: Design and experimental verification,'' \emph{IEEE
  Transactions on Smart Grid}, vol.~6, no.~2, pp. 618--626, 2015.

\bibitem{gelenbe2012energy}
E.~Gelenbe, ``Energy packet networks: adaptive energy management for the
  cloud,'' in \emph{Proceedings of the 2nd International Workshop on Cloud
  Computing Platforms}.\hskip 1em plus 0.5em minus 0.4em\relax ACM, 2012, p.~1.

\bibitem{gelenbe2016energy}
E.~Gelenbe and E.~T. Ceran, ``Energy packet networks with energy harvesting,''
  \emph{IEEE Access}, vol.~4, pp. 1321--1331, 2016.

\bibitem{gelenbe1994g}
E.~Gelenbe, ``G-networks: a unifying model for neural and queueing networks,''
  \emph{Annals of Operations Research}, vol.~48, no.~5, pp. 433--461, 1994.

\bibitem{rodriguez2015experimental}
Rodriguez-Bernuz \emph{et~al.}, ``Experimental validation of a single phase
  intelligent power router,'' \emph{Sustainable Energy, Grids and Networks},
  vol.~4, pp. 1--15, 2015.

\bibitem{lee2011demand}
S.~Lee, S.~Kim, and S.~Kim, ``Demand side management with air conditioner loads
  based on the queuing system model,'' \emph{IEEE Transactions on Power
  Systems}, vol.~26, no.~2, pp. 661--668, 2011.

\bibitem{zhang2013novel}
B.~Zhang and J.~Baillieul, ``A novel packet switching framework with binary
  information in demand side management,'' in \emph{Decision and Control (CDC),
  2013 IEEE 52nd Annual Conference on}.\hskip 1em plus 0.5em minus 0.4em\relax
  IEEE, 2013, pp. 4957--4963.

\bibitem{rezaei2014packetized}
P.~Rezaei, J.~Frolik, and P.~D. Hines, ``Packetized plug-in electric vehicle
  charge management,'' \emph{IEEE Transactions on Smart Grid}, vol.~5, no.~2,
  pp. 642--650, 2014.

\bibitem{frolik2018systems}
J.~Frolik, P.~Hines, and M.~Almassalkhi, ``Systems and methods for randomized,
  packet-based power management of conditionally-controlled loads and
  bi-directional distributed energy storage systems,'' Mar.~22 2018, uS Patent
  App. 15/712,089.

\bibitem{nardelli2017multi}
P.~H. Nardelli and F.~K{\"u}hnlenz, ``Multi-layer analysis of iot-based
  systems,'' \emph{arXiv preprint arXiv:1708.06506}, 2017.

\bibitem{nardelli2014models}
P.~H. Nardelli \emph{et~al.}, ``Models for the modern power grid,'' \emph{The
  European Physical Journal Special Topics}, vol. 223, no.~12, pp. 2423--2437,
  2014.

\bibitem{Wang2011}
\BIBentryALTinterwordspacing
W.~Wang, Y.~Xu, and M.~Khanna, ``{A survey on the communication architectures
  in smart grid},'' \emph{Computer Networks}, vol.~55, no.~15, pp. 3604--3629,
  Oct. 2011. [Online]. Available:
  \url{http://www.sciencedirect.com/science/article/pii/S138912861100260X}
\BIBentrySTDinterwordspacing

\bibitem{popovski2018wireless}
P.~Popovski, J.~J. Nielsen, C.~Stefanovic, E.~de~Carvalho, E.~Strom, K.~F.
  Trillingsgaard, A.-S. Bana, D.~M. Kim, R.~Kotaba, J.~Park \emph{et~al.},
  ``Wireless access for ultra-reliable low-latency communication: Principles
  and building blocks,'' \emph{IEEE Network}, vol.~32, no.~2, pp. 16--23, 2018.

\bibitem{liu2018sparse}
L.~Liu, E.~G. Larsson, W.~Yu, P.~Popovski, C.~Stefanovic, and E.~de~Carvalho,
  ``Sparse signal processing for grant-free massive connectivity: A future
  paradigm for random access protocols in the internet of things,'' \emph{IEEE
  Signal Processing Magazine}, vol.~35, no.~5, pp. 88--99, 2018.

\bibitem{Pinomaa2011}
A.~Pinomaa, J.~Ahola, and A.~Kosonen, ``Plc concept for lvdc distribution
  systems,'' \emph{IEEE Communications Magazine}, vol.~49, no.~12, pp. 55--63,
  Dec. 2011.

\bibitem{Kansal2012}
P.~Kansal and A.~Bose, ``Bandwidth and latency requirements for smart
  transmission grid applications,'' \emph{IEEE Transactions on Smart Grid},
  vol.~3, no.~2, pp. 1344--1352.

\bibitem{Noam2012}
E.~M. Noam, L.~M. Pupillo, and J.~J. Kranz, ``Broadband newtworks, smart grids,
  and climate change,'' in \emph{Springer Science and Business Media}.\hskip
  1em plus 0.5em minus 0.4em\relax Springer, 2012, p. 254.

\bibitem{long2018peer}
C.~Long, J.~Wu, Y.~Zhou, and N.~Jenkins, ``Peer-to-peer energy sharing through
  a two-stage aggregated battery control in a community microgrid,''
  \emph{Applied Energy}, vol. 226, pp. 261--276, 2018.

\bibitem{van2015power}
T.~Van Der~Schoor and B.~Scholtens, ``Power to the people: Local community
  initiatives and the transition to sustainable energy,'' \emph{Renewable and
  Sustainable Energy Reviews}, vol.~43, pp. 666--675, 2015.

\bibitem{kuhnlenz2018implementing}
F.~K{\"u}hnlenz, P.~H. Nardelli, S.~Karhinen, and R.~Svento, ``Implementing
  flexible demand: Real-time price vs. market integration,'' \emph{Energy},
  vol. 149, pp. 550--565, 2018.

\bibitem{Nokia2016}
Nokia, ``{5G Masterplan - five keys to create the new communications era},''
  Tech. Rep., 2016.

\bibitem{5GEnergy}
\BIBentryALTinterwordspacing
L.~Thrybom and {\'{A}}.~Kapovits, ``{5G and Energy},'' 5G-PPP, Tech. Rep.
  September, 2015. [Online]. Available:
  \url{https://5g-ppp.eu/wp-content/uploads/2014/02/5G-PPP-White{\_}Paper-on-Energy-Vertical-Sector.pdf}
\BIBentrySTDinterwordspacing

\bibitem{NokiacMTC2016}
Nokia, ``{5G for Mission Critical Communication:Achieve ultra-reliability and
  virtual zero latency},'' Tech. Rep., 2016.

\bibitem{Kuzlu2014}
\BIBentryALTinterwordspacing
M.~Kuzlu, M.~Pipattanasomporn, and S.~Rahman, ``{Communication network
  requirements for major smart grid applications in HAN, NAN and WAN},''
  \emph{Comput. Networks}, vol.~67, pp. 74--88, 2014. [Online]. Available:
  \url{http://dx.doi.org/10.1016/j.comnet.2014.03.029}
\BIBentrySTDinterwordspacing

\bibitem{8120238MMTC_LTE}
P.~Andres-Maldonado, P.~Ameigeiras, J.~Prados-Garzon, J.~Navarro-Ortiz, and
  J.~M. Lopez-Soler, ``Narrowband iot data transmission procedures for massive
  machine-type communications,'' \emph{IEEE Network}, vol.~31, no.~6, pp.
  8--15, November 2017.

\bibitem{Lopez2017}
\BIBentryALTinterwordspacing
O.~L.~A. L{\'{o}}pez \emph{et~al.}, ``{Aggregation and Resource Scheduling in
  Machine-type Communication Networks},'' pp. 1--33, Aug. 2017. [Online].
  Available: \url{http://arxiv.org/abs/1708.07691}
\BIBentrySTDinterwordspacing

\bibitem{Guo2017}
J.~Guo \emph{et~al.}, ``{Massive Machine Type Communication with Data
  Aggregation and Resource Scheduling},'' \emph{IEEE Trans. Commun.}, vol.
  6778, no.~c, pp. 1--1, 2017.

\bibitem{Shirvanimoghaddam2017b}
M.~Shirvanimoghaddam, M.~Dohler, and S.~J. Johnson, ``{Massive Non-Orthogonal
  Multiple Access for Cellular IoT: Potentials and Limitations},'' \emph{IEEE
  Commun. Mag.}, vol.~55, no.~9, pp. 55--61, 2017.

\bibitem{Durisi2016a}
G.~Durisi, T.~Koch, and P.~Popovski, ``{Toward Massive, Ultrareliable, and
  Low-Latency Wireless Communication With Short Packets},'' \emph{Proc. IEEE},
  vol. 104, no.~9, pp. 1711--1726, Sep. 2016.

\bibitem{Ozmen2016}
M.~Ozmen and M.~C. Gursoy, ``{Wireless Throughput and Energy Efficiency With
  Random Arrivals and Statistical Queuing Constraints},'' \emph{IEEE Trans.
  Inf. Theory}, vol.~62, no.~3, pp. 1375--1395, Mar. 2016.

\bibitem{giordano2012business}
V.~Giordano and G.~Fulli, ``A business case for smart grid technologies: A
  systemic perspective,'' \emph{Energy Policy}, vol.~40, pp. 252--259, 2012.

\bibitem{gregoratti2015distributed}
D.~Gregoratti and J.~Matamoros, ``Distributed energy trading: The
  multiple-microgrid case,'' \emph{iEEE Transactions on industrial
  Electronics}, vol.~62, no.~4, pp. 2551--2559, 2015.

\bibitem{huang2019embedding}
Y.~Huang, G.~Poderi, S.~{\v{S}}{\'c}epanovi{\'c}, H.~Hasselqvist, M.~Warnier,
  and F.~Brazier, ``Embedding internet-of-things in large-scale socio-technical
  systems: A community-oriented design in future smart grids,'' in \emph{The
  Internet of Things for Smart Urban Ecosystems}.\hskip 1em plus 0.5em minus
  0.4em\relax Springer, 2019, pp. 125--150.

\bibitem{pouttu2017p2p}
A.~Pouttu \emph{et~al.}, ``P2p model for distributed energy trading, grid
  control and ict for local smart grids,'' in \emph{Networks and Communications
  (EuCNC), 2017 European Conference on}.\hskip 1em plus 0.5em minus 0.4em\relax
  IEEE, 2017, pp. 1--6.

\bibitem{steg2018drives}
L.~Steg, R.~Shwom, and T.~Dietz, ``What drives energy consumers?: Engaging
  people in a sustainable energy transition,'' \emph{IEEE Power and Energy
  Magazine}, vol.~16, no.~1, pp. 20--28, 2018.

\bibitem{perlaviciute2018heart}
G.~Perlaviciute \emph{et~al.}, ``At the heart of a sustainable energy
  transition: The public acceptability of energy projects,'' \emph{IEEE Power
  and Energy Magazine}, vol.~16, no.~1, pp. 49--55, 2018.

\bibitem{ramezanipour2018increasing}
I.~Ramezanipour, P.~H. Nardelli, H.~Alves, and A.~Pouttu, ``Increasing the
  throughput of an unlicensed wireless network through retransmissions,'' in
  \emph{2018 IEEE 87th Vehicular Technology Conference (VTC Spring)}.\hskip 1em
  plus 0.5em minus 0.4em\relax IEEE, 2018, pp. 1--5.

\bibitem{ramezanipour2018finite}
I.~Ramezanipour, P.~Nouri, H.~Alves, P.~H. Nardelli, R.~D. Souza, and
  A.~Pouttu, ``Finite blocklength communications in smart grids for dynamic
  spectrum access and locally licensed scenarios,'' \emph{IEEE Sensors
  Journal}, vol.~18, no.~13, pp. 5610--5621, 2018.

\bibitem{alves2018secure}
H.~Alves, P.~H. Nardelli, and C.~H. de~Lima, ``Secure statistical qos
  provisioning for machine-type wireless communication networks,'' in
  \emph{2018 IEEE 87th Vehicular Technology Conference (VTC Spring)}.\hskip 1em
  plus 0.5em minus 0.4em\relax IEEE, 2018, pp. 1--5.

\bibitem{lopez2018aggregation}
O.~L.~A. L{\'o}pez, H.~Alves, P.~H.~J. Nardelli, and M.~Latva-aho,
  ``Aggregation and resource scheduling in machine-type communication networks:
  A stochastic geometry approach,'' \emph{IEEE Transactions on Wireless
  Communications}, 2018.

\bibitem{tome2018long}
M.~Tome, P.~Nardelli, and H.~Alves, ``Long-range low-power wireless networks
  and sampling strategies in electricity metering,'' \emph{IEEE Transactions on
  Industrial Electronics}, 2018.

\bibitem{tome2018event}
M.~C. Tome, P.~H.~J. Nardelli, and H.~Alves, ``Event-based electricity
  metering: An autonomous method to determine transmission thresholds,'' in
  \emph{2018 IEEE 87th Vehicular Technology Conference (VTC Spring)}.\hskip 1em
  plus 0.5em minus 0.4em\relax IEEE, 2018, pp. 1--5.

\bibitem{tome2018storage}
------, ``Event-based storage management in modern electricity power grids,''
  in \emph{4th International Conference on Event-Based Control, Communication,
  and Signal Processing (EBCCSP 2018)}, 2018.

\bibitem{kuhnlenz2018demand}
F.~Kühnlenz, P.~H.~J. Nardelli, and H.~Alves, ``Demand control management in
  microgrids: The impact of different policies and communication network
  topologies,'' \emph{IEEE Systems Journal}, pp. 1--8, 2018.

\bibitem{kuhnlenz2016dynamics}
F.~K{\"u}hnlenz and P.~H. Nardelli, ``Dynamics of complex systems built as
  coupled physical, communication and decision layers,'' \emph{PloS one},
  vol.~11, no.~1, p. e0145135, 2016.

\bibitem{nardelli2017smart}
P.~H.~J. Nardelli and F.~K{\"u}hnlenz, ``Why smart appliances may result in a
  stupid energy grid?'' \emph{IEEE Systems, Man, and Cybernetics Magazine},
  2018, to appear.

\bibitem{kuhnlenz2017agent}
F.~K{\"u}hnlenz and P.~H. Nardelli, ``Agent-based model for spot and balancing
  electricity markets,'' in \emph{Communications Workshops (ICC Workshops),
  2017 IEEE International Conference on}.\hskip 1em plus 0.5em minus
  0.4em\relax IEEE, 2017, pp. 1123--1127.

\bibitem{pilgerstorfer2017self}
P.~Pilgerstorfer and E.~Pournaras, ``Self-adaptive learning in decentralized
  combinatorial optimization: a design paradigm for sharing economies,'' in
  \emph{Proceedings of the 12th International Symposium on Software Engineering
  for Adaptive and Self-Managing Systems}.\hskip 1em plus 0.5em minus
  0.4em\relax IEEE Press, 2017, pp. 54--64.

\bibitem{pournaras2017self}
E.~Pournaras, M.~Yao, and D.~Helbing, ``Self-regulating supply--demand
  systems,'' \emph{Future Generation Computer Systems}, vol.~76, pp. 73--91,
  2017.

\bibitem{nambi2016temporal}
S.~Nambi, E.~Pournaras, and R.~V. Prasad, ``Temporal self-regulation of energy
  demand,'' \emph{IEEE Transactions on Industrial Informatics}, vol.~12, no.~3,
  pp. 1196--1205, 2016.

\bibitem{demidov2018fusion}
I.~Demidov, ``Fusion grid concept design and techno-economic analysis for
  developing countries,'' \emph{Master's thesis}, 2018, available at:
  \url{http://lutpub.lut.fi/handle/10024/158601}.

\bibitem{hosseini2016overview}
S.~A. Hosseini, H.~A. Abyaneh, S.~H.~H. Sadeghi, F.~Razavi, and A.~Nasiri, ``An
  overview of microgrid protection methods and the factors involved,''
  \emph{Renewable and Sustainable Energy Reviews}, vol.~64, pp. 174--186, 2016.

\end{thebibliography}

\end{document}